\documentstyle[preprint,aps,epsf]{revtex}
\begin{document}
\draft
\title{Small polaron formation in dangling-bond wires on the Si(001) surface}
\author{D.R.~Bowler\cite{drb} and A.J.~Fisher\cite{ajf}}
\address{Department of Physics and Astronomy, University College London,\\
Gower Street, London WC1E 6BT, UK}
\maketitle

\begin{abstract}
From electronic structure calculations, we find that carriers injected
into dangling-bond atomic wires on the Si(001) surface will self-trap
to form localised polaron states.  The self-trapping distortion takes
the form of a local suppression of the buckling of the dimers in the
wire, and is qualitatively different for the electron and hole
polarons.  This result points to the importance of polaronic effects
in understanding electronic motion in such nanostructures.
\end{abstract}


Developments in nanotechnology are placing increasing demands on our
understanding of electron transport in confined systems; this applies
particularly to coherent quantum transport and tunnelling phenomena.
Among the most promising one-dimensional structures that have been
produced by nanolithography are `dangling-bond wires', formed by the
selective removal of H atoms from the H-saturated Si(001) surface with
a scanning tunnelling microscope\cite{expts,firsttheory}.  In this
letter we show that carriers injected into these dangling-bond wires
self-localise to form polarons; this process is crucial in determining
the transport properties of the wires.

There is good experimental evidence that a Peierls distortion occurs
at low teperatures in a dangling-bond wire\cite{hitosugi99}, although there
is still debate about the extent of the underlying structural
distortion\cite{firsttheory,hitosugi99}; this distortion removes the
one-dimensional metallic density of states and produces a narrow-gap
one-dimensional semiconductor\cite{Peierls}.  In systems with Peierls
distortions, charge transport is dominated by polaron species
\cite{heeger88,fisher89}; this is true both when charge is permanently
introduced into the wire by doping, and also when the charge enters
the chain only transiently, as part of a quantum mechanical tunnelling
event \cite{ourprl}.  However, recent calculations of the transport
properties of dangling-bond wires do not include polaronic effects
\cite{doumergue99}.  It is therefore of great importance to determine
whether such polarons are, in fact, formed in this system, where the
electrons in the wire are strongly coupled to a three-dimensional
substrate and so a purely one-dimensional picture is not applicable.

In this letter we provide clear theoretical evidence that polarons
will form when holes (or, as discussed later, electrons) are injected
into a dangling-bond wire.  We calculate the binding energy, atomic
distortion, and wavefunction localisation of the polarons, and provide
realistic estimates of the effective mass and of the effect that
polaron formation will have on the coherent transport of charge along
the wire.

We use three different methods to perform our calculations.  The first
is density functional theory in the local density approximation (LDA),
with a plane-wave basis set and norm-conserving
pseudopotentials\cite{kerker,kleinman,payne}.  We use a plane-wave
energy cutoff of 200\,eV.  However, the formation of polarons is
expected to involve long-range atomic distortions extending over many
surface unit cells; to capture these effects, an accurate but
approximate method is required which includes the interplay between
electronic and atomic structure.  We therefore use two variants of a
total-energy orthogonal tight-binding method\cite{sutton88,goringe97},
implemented in the OXON package.  In one form the wavefunctions are
calculated by explicit diagonalisation of the tight-binding
Hamiltonian (we shall refer to this as ${\rm O}(N^3)$ tight binding,
since the computational effort scales as the cube of the system size).
In the other form, the one-electron density matrix is calculated
directly, while truncating it at a finite number of nearest-neighbour
`hops', to produce a scheme which scales linearly with system size
(${\rm O}(N)$ tight binding) \cite{li93,goringe95}.  The same
tight-binding parameters are used in both cases; the silicon-silicon
parameters\cite{chadi79,bowler98} have been shown to provide a good
account of the elastic properties and band structure of bulk Si, while
the silicon-hydrogen parameters \cite{bowler98} give a good account of
the structure and diffusion of H on
Si(001)\cite{owen96,bowler98b,bowler99}.  The ${\rm O}(N)$ method is
extremely efficient for searching configuration space in large unit
cells; all the structures and energies were subsequently validated
with the ${\rm O}(N^3)$ technique, and it is these numbers that are
reported here.

\begin{figure}[ht]

\begin{tabular}{c}
\epsfxsize=100mm
\epsfbox{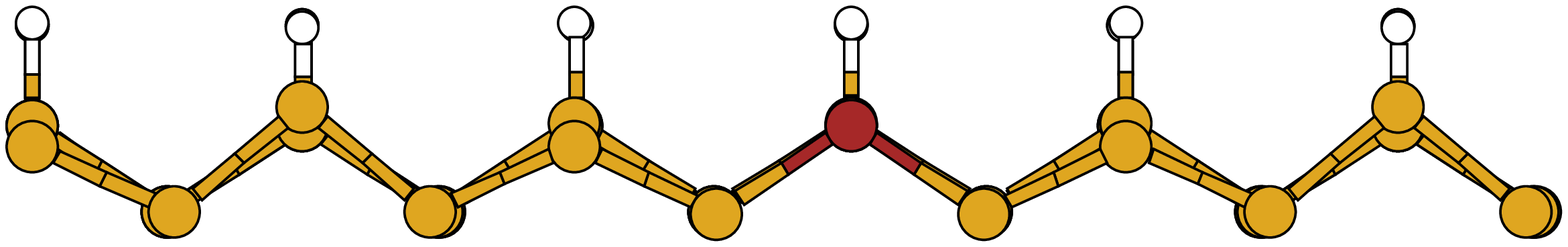}\\
(a)\\
\epsfxsize=100mm
\epsfbox{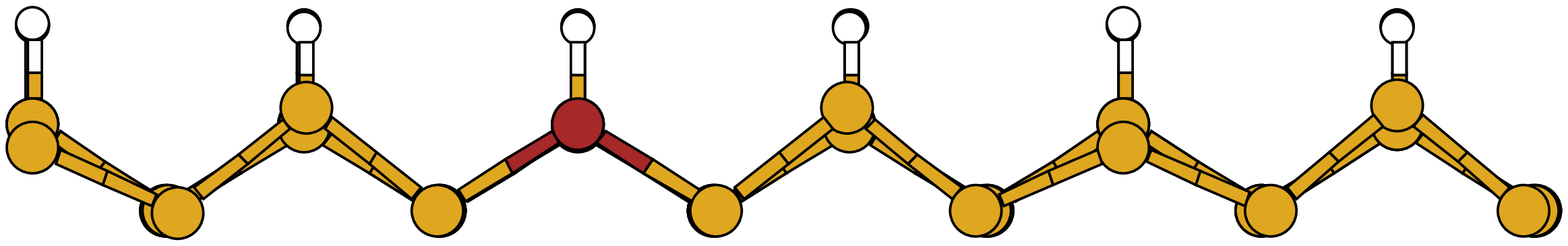}\\
(b)\\
\end{tabular}

\caption{\label{structure} Part of the geometry used in the
calculations.  Only the topmost atoms are shown in (a) the hole
polaron and (b) the electron polaron calculations.  The small white
circles are H; light grey circles are Si; the Si atom on which the
polaron traps is darkened in both cases.  The alternation between `up'
and `down' atoms along the wire, interrupted by the formation of the
polaron, is visible in the figures.  The actual length of the unit
cell was 12 dimers in each case}
\end{figure}

We performed calculations on several different unit cells, always
using a slab geometry with periodic boundary conditions.  The number
of Si layers perpendicular to the surface is 5 for the LDA
calculations (with the final layer terminated in H and constrained to
lie in bulk-like positions), 10 for the ${\rm O}(N)$ tight-binding
calculations (with the final layer terminated in H, and the bottom
five layers constrained to lie in bulk-like positions), and 6 for the
${\rm O}(N^3)$ tight-binding (again H-terminated, with the bottom
layer frozen).  The lengths of the different cells along the wire axis
were 2 and 12 dimers, and the cells were two dimer rows wide in all
cases.  A single dangling-bond wire was introduced into the
calculations by removing one line of H atoms along a dimer row
direction.  The topmost section of the cell  is shown in
Figure~\ref{structure}, with the trapping distortions of the hole and
electron polarons.

We begin by describing our calculations of the wire in its uncharged
state.  We used the 2-dimer cell, and performed the calculations using
both the LDA and tight-binding (using a $2\times2\times2$
Monkhorst-Pack $k$-point mesh in both cases, making use of
time-reversal symmetry); this gave us a check on the accuracy of the
tight-binding scheme with which we subsequently treated the polaron
species.  In agreement with previous work, we found that the Peierls
distortion manifested itself as an alternating up and down
displacement of the clean Si atoms forming the dangling bond wire.
Table~\ref{bonds} shows the key structural parameters obtained in LDA
and by tight-binding in the 2-dimer cell; the only significant
difference is in the Si-H bond lengths, which are not important for
the distortions we discuss in this paper.  The height difference
$\Delta z$ between the `up' and `down' Si atoms in the wire is well
reproduced.  The relaxation energies found from both LDA and
tight-binding calculations agree well.

\begin{figure}[ht]

\begin{tabular}{c}
\protect 
\epsfxsize=80mm
\epsfbox{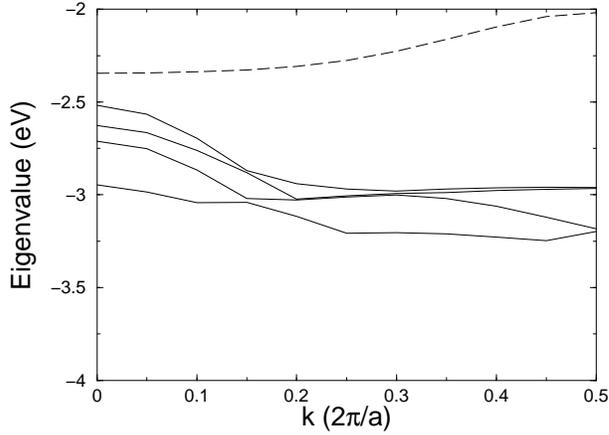}\\
(a)\\
\epsfxsize=80mm
\epsfbox{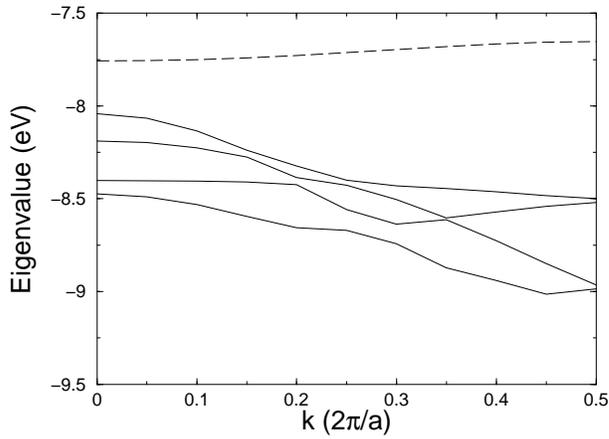}\\
(b)\\
\end{tabular}

\caption{\label{bands} The bandstructure near the Fermi energy of an
uncharged dangling-bond wire: (a) Kohn-Sham eigenvalues calculated in
the LDA; (b) eigenvalues from ${\rm O}(N^3)$ tight-binding
calculations in which the Hamiltonian is explicitly diagonalised.  The
five highest occupied bands are shown; the relatively flat highest
band (shown dashed) is predominantly localised on the dangling bonds.}
\end{figure}

The electron density is higher on the up atom (reflecting the move
towards sp$^3$ hybridisation) and lower on the down atom (which is
more sp$^2$-like, with an empty p orbital).  This is somewhat
analogous to the tilted dimer structure seen on the clean Si(001)
surface.  The band structures for the highest occupied states from the
LDA and from the ${\rm O}(N^3)$ tight-binding are compared in
Figure~\ref{bands}.  We see that the two are very similar; in
particular, the dispersion of the highest occupied band is very
similar in the two cases.  This is important, because the electronic
states associated with the polaron are predominantly derived from this
bands.  In the course of this comparison, we found that the structural
and electronic properties of the system are very sensitive to the
surface $k$-point sampling used to construct the LDA charge density.
Relaxation at a single special $k$-point in the surface Brillouin zone
gave a non-zero Peierls distortion, but produced an electronic
structure with qualitatively different dispersion; however, a
$2\times2\times2$ Monkhorst-Pack mesh was found to be sufficient.
This is not surprising for a system exhibiting a Peierls distortion,
where the charge density is expected to be sensitive to the $k$-point
sampling in the vicinity of the Fermi points.

\begin{figure}

\begin{center}
\epsfxsize=80mm
\epsfbox{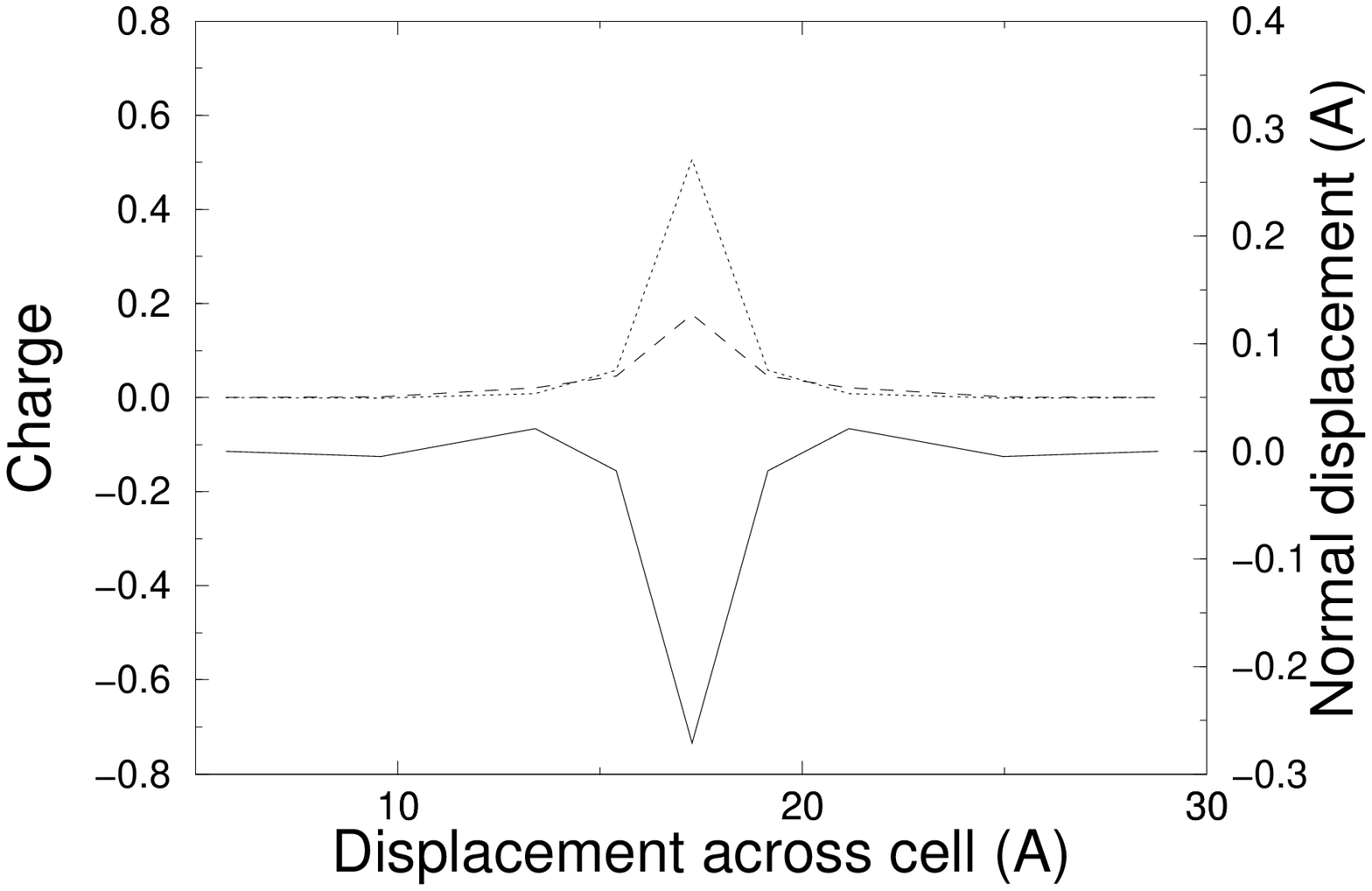}
\end{center}

\caption{\label{hdistortion} Changes in local atomic properties when a
hole polaron is formed in a dangling-bond wire.   The atoms shown are
those Si atoms most strongly affected: these are those bare atoms in
the topmost line, and the two second-layer neighbours of the polaron
trapping site.  Displacements of the
wire atoms normal to the surface as a function of distance along the
wire (solid curve, right-hand scale); distribution of the excess
charge among the same atoms (dotted line, left-hand scale); charge
density of the highest occupied electronic state (dashed line,
left-hand scale).}

\end{figure}

We investigated the formation of a hole polaron using tight binding by
removing a single electron from the 12-dimer cell.  This cell contains
612 atoms (for the ${\rm O}(N)$ method) or 420 atoms (${\rm O}(N^3)$
method); bulk-like atoms were removed from the bottom of the slab for
the ${\rm O}(N^3)$ calculations for reasons of computational
efficiency, with negligible effect on the results.  Only the
$\Gamma$-point of the reduced Brillouin zone was sampled.  A single
`up' atom of the wire was displaced downwards normal to the surface to
break the translational symmetry, and the geometry was then
relaxed. It did not revert to the original translationally-ordered
state; instead, a localised pattern of distortions occurred in the Si
atoms of the wire, predominantly in the direction normal to the
surface (Figure~\ref{hdistortion}).  (We also found that the symmetry
of the translationally-ordered state spontaneously broke to form the
same structure.)  This distortion was associated with a localisation
of the charge. The relaxation energy of the charged system from its
translationally ordered state into the polaron state was 0.21\,eV.
The highest occupied state is found to be localised in the region of
the distortion.

\begin{figure}%

\begin{center}
\epsfxsize=80mm
\epsfbox{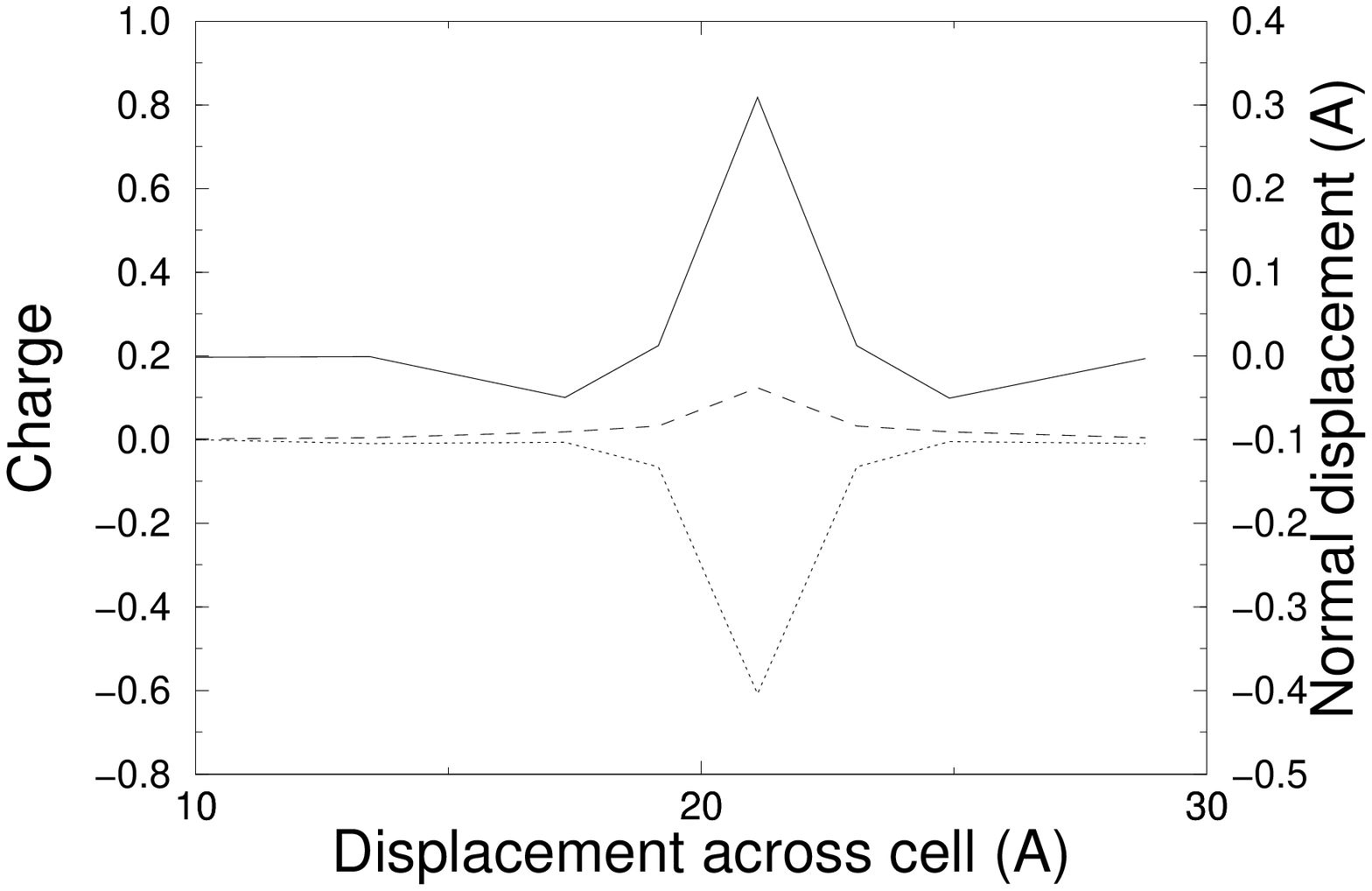}
\end{center}

\caption{\label{edistortion}Changes in local atomic properties when an
electron polaron is formed in a dangling-bond wire.  The atoms shown
are those Si atoms most strongly affected: these are those bare atoms
in the topmost line, and the two second-layer neighbours of the
polaron trapping site.  Displacements of the wire atoms normal to the
surface as a function of distance along the wire (solid curve,
right-hand scale); distribution of the excess charge among the same
atoms (dotted line, left-hand scale); charge density of the highest
occupied electronic state (dashed line, left-hand scale).  Note that
the excess charge is now negative, while the charge density remains
positive.}

\end{figure}

We also calculated the relaxation that follows the addition of an
electron to the wire.  Once again, a polaron is formed; the major
effect now is that one of the `down' wire atoms moves up, accompanied
by some distortion of the neighbouring atoms
(Figure~\ref{edistortion}).  Once again, a localised distribution of
the excess charge and a localised one-electron state are produced.
The binding energy of the electron polaron in the 12-dimer cell is
0.16\,eV.  This is entirely reasonable and consistent with what is
predicted for the hole; however, since our tight-binding has been
fitted to structural (rather than excited-state) properties and so may
not reproduce empty states as well as filled states, we concentrate on
the results for the hole polaron in the remainder of the paper.

It is straightforward to understand qualitatively the driving forces
behind these displacements.  When an electron is removed, there is a
tendency to drive the last (now half-occupied) filled state upwards in
energy; this can be accomplished by driving one of the sp$^3$-like
`up' atoms downwards and making it more sp$^2$-like.  Conversely, when
an electron is added, there is an advantage in lowering the energy of
a formerly empty state; this can be achieved by moving an sp$^2$-like `down'
atom upwards to make it more sp$^3$-like.  Both motions correspond to
a local reduction of the Peierls distortion in a manner appropriate to
the sign of the added charge.

It is interesting to compare these polarons with those formed in
conducting polymers\cite{heeger88}.  In polymers, the dominant
distortion is in the bond lengths, and the charge-conjugation symmetry
in the system (which holds exactly in the simple Su-Schrieffer-Heeger
model\cite{ssh} and approximately in the real molecules) implies that the
electron and hole polarons are similar.  In the dangling-bond wire, on
the other hand, the dominant distortions are perpendicular to the
surface, and the electron and hole polarons are qualitatively
different.  In both cases, however, one can think of the polaron
distortion as producing a local suppression of the band gap; in the
conducting polymer this arises from a local reduction of the
bond-length alternation, whereas in the dangling-bond wire it arises
from the vertical displacement of Si atoms.  A further difference is
that the distortion in the dangling-bond is more localized; while this
makes it particularly easy to justify it {\em post hoc}, its form
could not have been predicted before our calculations.

In order to quantify the coupling of the carriers to the lattice, and
to estimate the dynamical properties of the polarons, we have analysed
the polaron distortion in terms of the phonon spectrum of the neutral
wire.  First we check to see whether the harmonic approximation
correctly reproduces the energy change when a neutral system is
distorted to the geometry of the hole polaron; we find the
harmonic approximation gives $\Delta E=0.39\,\rm eV$, in fair
agreement with the full tight-binding result of $\Delta E=0.30\,\rm
eV$.  (The dominant source of the anharmonicity is in the softening of
the flattening Si dimers as the polaron is formed.  We have checked
that the harmonic approximation becomes exact as the magnitude of the
distortion is reduced.)  Now for each mode $q$ we can calculate a
Huang-Rhys factor $S_q=\Delta E_q/(\hbar\omega_q)$, where $\Delta E_q$
is the relaxation energy of mode $q$ in the process of forming the
polaron and $\omega_q$ is the frequency\cite{amsbook}.  We find that
$S=\sum_qS_q=0.125$, putting the system in the $S<1$ weak-coupling
regime; we also find that the maximum value of an individual $S_q$ is
0.0056, showing that there is no single mode dominating the distortion.
As might be expected, the modes with the largest $S_q$ are those
corresponding to motions of the depassivated Si atoms normal to the
surface.

We can also find the Huang-Rhys factors $S'_q$ corresponding to the
displacement of the polaron by two dimers along the wire.  We take two
adjacent positions for the polaron and calculate a new set of $S'_q$
from the differences in the phonon displacements.  We find now that
$S'=\sum_qS'_q=0.251$---this is very close to $2S$, which can be
understood because the displacements from the uncharged state
corresponding to the two different polaron positions are almost
orthogonal in configuration space.  Using this result, and the bare
electronic bandstructure of the wire, we can estimate the
low-temperature effective mass of the polaronic species: the effective
hopping integral for the carrier between neighbouring sites is reduced
by approximately $\exp(-\sum_qS_q')$ \cite{holstein}, leading to an
enhancement in the effective mass from the bare hole band mass
$m^*=3.30\,m_e$ (computed from the tight-binding bandstructure) to
$m^*=5.45\,m_e$ for low-temperature coherent transport.  Simulations
of the classical diffusion of the polarons at high temperatures will
be presented in a future publication.

This enables us to assess the likely effect of the polaron formation
on electron transport in the wire.  If the chemical potential is close
to the Fermi level of the substrate, transport through the wire will
be by coherent tunnelling.  In this regime, strong enhancement of the
tunnelling at low temperatures is to be expected, as found in
\cite{ourprl}.  There will be a crossover to thermal excitation of
polarons as the temperature rises.  If, on the other hand, the wire is
subject to such a strong departure from equilibrium that real (as
opposed to virtual) carriers can be injected into it, their effective
mass will be enhanced at low temperatures.  For both tunnelling and
band transport, we can estimate the crossover temperature to
thermally-activated hopping motion as being of the order of half the
frequency of the most strongly coupled modes \cite{langfirsov}; hence,
we find $T\approx 190\,\rm K$ for the hole, and about 30\% higher for
the electron.  We therefore expect that room-temperature experiments
will be in the high-temperature regime, but that the low-temperature
limit should also be experimentally accessible.

\acknowledgements We thank the UK Engineering and Physical Sciences
Research Council for the award of an Advanced Fellowship (AJF) and a
Postdoctoral Fellowship in Theoretical Physics (DRB), and for support
under grants GR/M09193 and GR/M01753.  We are grateful to Dr H.~Ness
and Professor A.M.~Stoneham for a number of discussions.

\newpage

\newpage

\newpage

\newpage

\begin{table}[ht]
\begin{tabular}{lcc}
Bond length (\AA)&LDA&Tight binding\\
\hline
Si-Si in `up' wire dimer&2.42&2.43\\
Si-Si in `down' wire dimer&2.38&2.36\\
Si-Si in non-wire dimer&2.40&2.37\\
Si-H on `up' wire dimer&1.55&1.50\\
Si-H on `down' wire dimer&1.54&1.49\\
\hline
$\Delta z$ between `up' and `down' wire atoms (\AA)&0.67&0.57\\
Peierls energy (eV, per dimer pair)&0.21&0.23\\
\end{tabular}
\caption{\label{bonds}
Comparison of structural parameters for the neutral, translationally ordered,
dangling-bond wire from LDA and tight-binding calculations.}
\end{table}


\begin{thebibliography}{99}

\bibitem[*]{drb}Email david.bowler@ucl.ac.uk.

\bibitem[+]{ajf}Email andrew.fisher@ucl.ac.uk.

\bibitem{expts}T.-C.~Shen {\it et al.} Science {\bf 253} 1590 (1995).

\bibitem{firsttheory}S.~Watanabe, Y.A.~Ono, T.~Hashizume and Y.~Wada,
Phys.\ Rev.\ B {\bf 54} R17308 (1996); Surf.\ Sci.\ {\bf 386} 340 (1997).

\bibitem{Peierls}R.E.~Peierls {\it Quantum Theory of Solids\/}
(Clarendon Press, Oxford 1955) p110.

\bibitem{hitosugi99}T.~Hitosugi {\it et al.}, Phys.\ Rev.\ Lett.\ {\bf
82} 4034 (1999).

\bibitem{heeger88}A.J.~Heeger, S.~Kivelson, J.R.~Schrieffer and
W.-P.~Su, Rev.\ Mod.\ Phys.\ {\bf 60} 781 (1988).

\bibitem{fisher89}A.J.~Fisher, W.~Hayes and D.S.~Wallace, J.\ Phys.:
Cond.\ Matt.\ {\bf 1} 5567 (1989).

\bibitem{ourprl}H.~Ness and A.J.~Fisher, Phys.\ Rev.\ Lett.\ {\bf 83}
452 (1999).

\bibitem{doumergue99}P.~Doumergue {\it et al.}, Phys.\ Rev.\ B {\bf
59} 15910 (1999).

\bibitem{kerker}G.~Kerker, J.\ Phys.\ C {\bf 13} 189 (1980).

\bibitem{kleinman}L.~Kleinman and D.M.~Bylander, Phys.\ Rev.\ Lett. 
{\bf 48} 1425 (1982).

\bibitem{payne}CASTEP 4.2 Academic Version licensed under the UKCP-MSI
agreement (1999); M.C.~Payne {\it et al.}, Rev. Mod. Phys. {\bf 64} 1045 (1992).

\bibitem{sutton88}A.P.~Sutton, M.W.~Finnis, D.G.~Pettifor and
Y.~Ohta, J.\ Phys.\ C {\bf 21} 35 (1988).

\bibitem{goringe97}C.M.~Goringe, D.R.~Bowler and E.H.~Hern\'andez,
Rep.\ Prog.\ Phys. {\bf 60} 1447 (1997).

\bibitem{li93}X.P.~Li, W.~Nunes and D,~Vanderbilt, Phys.\ Rev.\ B {\bf
47} 10891 (1993).

\bibitem{goringe95}C.M.~Goringe, DPhil Thesis, University of Oxford (1995).

\bibitem{chadi79}D.J.~Chadi, J.\ Vac.\ Sci.\ Tech.\ {\bf 16} 1290 (1979).

\bibitem{bowler98}D.R.~Bowler {\it et al.}, J.\ Phys.: Condens.\
Matter {\bf 10} 3719 (1998).

\bibitem{owen96}J.H.G.~Owen {\it et al.}, Phys.\ Rev.\ B {\bf 54}
14153 (1996).

\bibitem{bowler98b}D.R.~Bowler, J.H.G.~Owen, K.~Miki and
G.A.D.~Briggs, Phys.\ Rev.\ B {\bf 57} 8790 (1998).

\bibitem{bowler99}D.R.~Bowler {\it et al.}, J.\ Phys.: Condens. Matter
(in press).

\bibitem{ssh}W.-P.~Su, J.R.~Schrieffer and A.J.~Heeger, Phys.\ Rev.\
Lett.\ {\bf 42} 1698 (1979).

\bibitem{amsbook}A.M.~Stoneham, {\it The Theory of Defects in Solids},
Oxford (1975).

\bibitem{holstein}T.~Holstein, Ann.\ Phys.\ {\bf 8} 325 and 343 (1959).

\bibitem{langfirsov}I.G.~Lang and Y.A.~Firsov, Sov.\ Phys.\ JETP {\bf 16} 
1301 (1963).

\end{thebibliography}
\end{document}